\documentstyle[11pt]{article}
\begin{document}

\begin{flushright}
hep-lat/9606002 \\
Liverpool Preprint: LTH 373\\
Helsinki Preprint: HU-TFT-96-21 \\
3rd June, 1996\\
\end{flushright}

\begin{center}
{\LARGE\bf
Non-perturbative determination of beta-functions and excited string 
states from lattices.
}\\[5mm]

{\bf  C. Michael\footnotemark[1] }\\
{Theoretical Physics Division, Dept. of Mathematical Sciences, 
University of Liverpool, Liverpool,  L69 3BX, U.K.}\\
{\bf and}\\
{\bf A.M.~Green\footnotemark[2], P.S.~Spencer\footnotemark[3]}\\
{Research Institute for Theoretical Physics, P.O. Box 9, FIN-00014 
University of Helsinki, Finland}\\

\end{center}
\setcounter{footnote}{1}
\footnotetext{email: {\tt cmi@liv.ac.uk}} 
\setcounter{footnote}{2}
\footnotetext{email: {\tt GREEN@phcu.helsinki.fi}}
\setcounter{footnote}{3}
\footnotetext{email: {\tt spencer@rock.helsinki.fi}}

\begin{abstract}

We use lattice sum rules for the static quark potential to determine the
beta-function for  symmetric and asymmetric lattices non-perturbatively.
We also study the colour field distributions in excited gluonic states. 

\end{abstract}

\section{Introduction}

In a lattice calculation, the observables are extracted as dimensionless
ratios  involving the lattice spacing $a$. It is important to obtain
results for a range  of $\beta$ (the bare lattice coupling coefficient
given for $SU(N)$ by $2N/g^2$) in order  to explore the approach to the
continuum limit $a \to 0$. The region of small lattice corrections is
reached  when the lattice spacing $a$ obtained from different
observables has a common $\beta$-dependence. Usually  this has been
explored by simulating at two different values of $\beta$ and matching
the results. Another route to this information is from   the lattice sum
rule approach which allows~\cite{mich,rothe} a derivative of a lattice
observable  with respect to $\beta$ to be evaluated at a fixed
$\beta$-value by  summing plaquette contributions over all space.
Expressing this in terms of a  lattice spacing $a$ determined from that
observable,  the $\beta$-function $b=d\beta/d \ln a$ can then be
extracted~\cite{gl,wup}. Moreover, a generalisation to a lattice with
different spacings $a_i$ in the  four directions coming from a
generalised Wilson action of the form $\beta_{ij} \Box_{ij}$ gives
access~\cite{cm} to derivatives with respect to these  coefficients
$\beta_{ij}$ by summing plaquettes  in the $ij$ plane. It is thus
possible to evaluate generalised  $\beta$-functions 
 \begin{equation}
{\partial \beta_{ij} \over \partial \ln a_k}=
S \ \hbox{if}\ k=i \ \hbox{or}\ j
\ \ \hbox{and} \ \
{\partial \beta_{ij} \over \partial \ln a_k}=
U \ \hbox{if}\ k\ne i \ \hbox{or}\ j
\end{equation}
 by working on a symmetric lattice ($a_i=a$) at a fixed $\beta$ value.  

For future comparison, the perturbative series for these quantities 
in terms of the bare lattice coupling $\alpha=g^2/4\pi=1/\pi\beta$ 
for $SU(2)$ colour fields are~\cite{karsch}
 \begin{equation}
 b=2(S+U)=-0.3715(1+0.49 \alpha+\dots ) 
\ \ \ \ \ \ \ U-S = 2\beta f = 2 \beta(1-1.13\alpha+ \dots) \ .
  \end{equation}
 \noindent Here we use lattice sum rule  techniques to calculate both
$b$ and $f$  non-perturbatively.

 We shall apply this sum rule technique to the potential between a
static  quark and antiquark at separation $R$.  As well as a sum rule
study to obtain the generalised $\beta$-functions from  the ground state
potential, we also  explore excited gluonic potentials -- both to
determine the generalised  $\beta$-functions and to compare with excited
string models of the  spatial distribution of the colour  fields
responsible for the gluonic excitation.

\section{Sum rules for potentials}

We consider the static quark potential for separation $R$ which 
is defined as $V(R)$ in lattice units.  Then the colour fields 
associated with this pair of static quarks can be measured using 
plaquettes of appropriate orientation. Sum rules 
have been derived to relate the sum over all spatial positions 
of these colour fields to $V(R)$ and its derivative~\cite{cm}:   

\begin{equation}
V+R {\partial V \over \partial R}= 
b \sum  ( {\cal E}_L + 2  {\cal E}_T  + 2  {\cal B}_T + {\cal B}_L) 
\label{TASU}
\end{equation} 
\begin{equation}
V+R {\partial V \over \partial R}= 
-4\beta f \sum  ( {\cal E}_L - {\cal B}_L) 
\label{TELSU}
\end{equation} 
\begin{equation}
V-R {\partial V \over \partial R}= 
- 4 \beta f \sum  (  {\cal E}_T  -   {\cal B}_T ) \ .
\label{TEPSU}
\end{equation}

Here $L$ refers to longitudinal and $T$ to transverse with respect to
the interquark separation axis. The notation is that $ {\cal E} $ and
${\cal B}$ refer to  the difference of the expectation values of the
plaquette in the appropriate orientation  in the presence of the static
quarks and  in the vacuum:  $ {\cal E}_{i}= \langle R| \Box_{i0}
|R\rangle - \langle 0| \Box_{i0} |0\rangle$, etc., with $\Box={1 \over
2} {\rm Tr} (1-U_{\Box})$.  Thus  they have the interpretation of  gauge
invariant averages of the fluctuation of squared  strengths of the
colour electric (${\cal E}$) and magnetic (${\cal B}$) fields. Since
they are a difference between the  value in the presence of the
(generalised) Wilson loop representing  the static quarks and the vacuum
value, either sign is possible.

Because the combination ${\cal E} \pm {\cal B}$ corresponds to 
action/energy, we refer to these sum rules as `action', `longitudinal 
energy' and `transverse energy' respectively.

The sum rules are derived~\cite{cm} for torelons where there is no self
energy associated with the  static quarks. For the more practical case 
of the interquark potential between static sources, a  self energy  term
must be included in each of the sum rules. This  self energy will be
independent of the inter-quark separation $R$  and hence can be removed
exactly by considering differences of  the sum rules for two different
$R$-values.

The remaining obstacle to an  evaluation of the sum rules  is the
presence of the derivative $dV/dR$. This arises in the sum rule 
derivation from  re-expressing the derivative $dV(R)/d\beta$ at fixed
number  of lattice spacings $R$. In the limit of small lattice spacing
$a$ and large number of lattice spacings $R$, this derivative can be
accurately expressed in terms of $dV(R)/dR$ at fixed $\beta$. For small
$R$,  however, there may be substantial systematic error from relating 
the $\beta$-dependence at fixed $R$ to the $R$-dependence at fixed
$\beta$. Thus we prefer to  retain accuracy by eliminating $dV/dR$ from
the sum rules. This leads to the relations  which can be used to obtain
$f$ and $b$:
 \begin{equation}
V(R_1)-V(R_2)=- 2 \beta f \sum 
( {\cal E}_L - {\cal B}_L + {\cal E}_T  -   {\cal B}_T )_{R_1} 
-\sum ( {\cal E}_L - {\cal B}_L + {\cal E}_T  -   {\cal B}_T )_{R_2}
\label{EF}
\end{equation} 
\begin{equation}
b 
={ 2 ( V(R_1)-V(R_2)) \left( 1+{ 
 \sum ({\cal E}_T  -   {\cal B}_T)_{R_1}-
 \sum ( {\cal E}_T  -   {\cal B}_T )_{R_2}
 \over 
 \sum ({\cal E}_L  -   {\cal B}_L)_{R_1}-
 \sum ( {\cal E}_L  -   {\cal B}_L )_{R_2}
}
\right)^{-1}
\over
 \sum  ( {\cal E}_L + 2  {\cal E}_T  + 2  {\cal B}_T + {\cal B}_L)_{R_1}
 -\sum ( {\cal E}_L + 2  {\cal E}_T  + 2  {\cal B}_T + {\cal B}_L)_{R_2}
} \ .
\label{AB}
\end{equation}

As well as using these sum rules for different values of $R$ to
eliminate the self energy terms, it is also possible to use the
difference between  excited states and ground states. The derivation of
the sum rules can be  extended to the case of an excited state and the
expressions are the same  as those given in Eqs.~(\ref{TASU}),
(\ref{TELSU}) and (\ref{TEPSU}). The self energy is associated  with the
static quarks themselves and  is also the same for excited  states and
ground states and so cancels fully for differences of energies between
excited and ground states.  In practice we use the E$_u$ representation
of D$_{4h}$ to provide the most  accessible excited state on a lattice.
For more details on these excited  potentials see ref~\cite{hyb,pm}.

\section{Lattice evaluation}

We use a $16^3 \times 32$ lattice at $\beta=2.4$ for this preliminary
study with the $SU(2)$ gauge group. For many purposes $SU(2)$ colour 
provides an excellent test bed for QCD studies. In this work the scale 
as set by the string tension corresponds to a lattice spacing $a=0.6$
GeV$^{-1}$.  Generalised Wilson loops of size $R \times T$ are evaluated
with the  spatial paths of length $R$ at $t=0$ and $T$ being
constructed~\cite{phm} from fuzzed links (using 2 and 13 recursive
iterations of the projected sum of \mbox{4 $\times$ straight link $+$ 4
U-bends}). We study the plaquette signal for all plaquette orientations 
at  time slices $t=T/2$. The correlation of this plaquette sum with the
Wilson  loop then gives the required  squared fluctuation of the colour
field. 

For the sum rule study, we need the sum of plaquettes over all space in
principle. On the lattice, we sum the plaquettes  over all transverse 
space (a $16 \times 16 $ plane) for each $r_L$. The sum over $r_L$ is
then approximated by selecting the region within $\pm2$  lattice units
of the Wilson loop (ie from $r_L= - 2$ to $R+2$). We checked that this
approximation  did not introduce  any significant error since the
missing $r_L$ region contributes noise  and no signal to the
correlation. The data sample used comes from a comprehensive study of 
60 blocks of 125 measurements separated by 4 update sweeps (3 
over-relaxation plus  one heatbath). Error estimates use a full bootstrap
analysis of these blocks.

The fuzzed link operator to create a static quark and antiquark at
separation $R$ with colour field in a state of a given symmetry will
have an expansion in terms of the  eigenstates of the transfer
matrix
 \begin{equation}
|R\rangle=c_0|V_0\rangle+c_1|V_1\rangle+\dots
 \end{equation}
 The required ground state contribution is enhanced as $T \to \infty$
but  at the expense of increasing noise.  Using a variational basis of 2
different fuzzing levels,  we were able to construct  very good  ground
state operators. This is essential  since the required ground state
correlation $\langle V_0|\Box|V_0\rangle$ is extracted  from lattice
observables using (for $t=T/2$)
 \begin{equation}
\langle R_0| \Box_t  |R_T\rangle = 
c_0^2 e^{-V_0 T} \left( \langle V_0|\Box|V_0\rangle + 
2 h \langle V_1 | \Box | V_0\rangle +  \dots \right)
 \end{equation}
with 
 \begin{equation}
h={ c_1 \over c_0} e^{-(V_1-V_0)t}\ .
 \end{equation}
 We can estimate the excited state coefficient $h$ from the study of 
the Wilson loop itself since
 \begin{equation}
W(T)=\langle R_0|R_T\rangle= c_0^2 e^{-V_0 T} \left( 1+ h^2 + \dots \right) 
 \end{equation}
 so that $h^2 \approx e^{-V_0(N-T)} W(T)/W(N)-1$, where $N$ is the
largest  time separation which is accurately measured  and $V_0 $ is the
best estimate of the ground state energy.  To limit the  contamination
of the plaquette correlation by excited states to less than 10\%, we
require $h < 0.05$ which implies  $h^2 < 0.0025$. For $T=2$, we obtain 
$h^2=0.0005$ for $R=2$,  for $R=4$ we find  $h^2=0.0025$, while for
$R=6$ we use three fuzzing levels (40, 16 and 0) as basis to  construct
a ground state with $h^2=0.0036$. Thus in each case, to reduce the
excited state contamination to 
{\mbox{\footnotesize{$\stackrel{<}{\sim}$}}} 10 \% , a gap of $t=T/2=1$
is sufficient  between the operators and the plaquette insertion. For 
electric plaquettes, their unit extension in the time direction implies
that  a total length of $T=3$ is needed. Our results for $T>3$ have
larger statistical errors  than those for $T=3$ but are completely
consistent with them within  these errors - so confirming  the above
analysis. Our results are collected in  Table~1 (using $T \ge 3$).

\begin{table}[ht]
\begin{center}
\begin{tabular}{cccll}
$R_2,\ R_1$ &T& $V(R_2)-V(R_1)$ &\ \  $-b$ &\ \  $f$ \\ \hline
2,\ 1 &3& 0.1892(1) & 0.502(9) & 0.64(2) \\  
3,\ 2 &3& 0.1188(2) & 0.357(19) & 0.62(3) \\  
&4&  & 0.307(32) & 0.64(8) \\  
&5&  & 0.315(64) & 0.64(15) \\  
4,\ 2 &3& 0.2129(3) & 0.362(10) & 0.59(3) \\ 
 &4&  & 0.349(23) & 0.64(7) \\ 
6,\ 4 &3& 0.1629(5) & 0.351(25) & 0.61(9) \\  
\end{tabular}
\caption{Generalised $\beta$-functions determined from sum rules.}
\end{center}
\end{table}

The lattice observables in the gauge sector with the Wilson action are 
known to approach the continuum limit as $a \to 0$ with corrections of 
order $a^2$. These corrections, sometimes called lattice artefacts, will
make the determination of the lattice spacing $a$ different for
different  physical observables on a lattice. The region of $a$ where
such differences are  small is called the scaling region and we expect
$\beta=2.4$ to be in that region.  Nevertheless, $d\beta/ d \ln a$ may
give different values when different observables  are used to define $a$
--  and that is one of the goals of this study. A particularly severe
manifestation arises with  static potentials at  small separation $R$ --
since the lattice artefacts are known to  behave as $a^2/R^2$ and are
important for $R \approx a$.  Indeed we see in  Table~1 that there are 
significant lattice artefacts  at $R=1$ which provide an explanation of 
the discrepancy in the first row of Table~1. We see consistent results
when $R>1$. This provides evidence  that, to our precision, 
non-perturbative determinations of the $\beta$-functions are independent
of  observable for potentials with $R>1$ at $\beta=2.4$.

The bare value of  $\alpha$ is $0.13$, which gives next to leading order
perturbative values of $b=-0.396$ and $f=0.85$. The  effective coupling
is expected to be approximately  twice as big which will decrease  the
perturbative estimate for $f$ and improve the  agreement with our
non-perturbative result. For $b$, however, such an  increase in
$\alpha$  will make the agreement even worse. Thus a  perturbative
evaluation of $b$ is unreliable.

The $\beta$-function has also been studied non-perturbatively on a
lattice  by matching the measured potentials at two $\beta$ values.
Between $\beta=2.4$ and 2.5, this gave~\cite{phm} $b=-0.277(4)$. The
value determined here is the derivative at $\beta=2.4$ rather  than
coming from a finite difference. The qualitative features of the two
approaches are in agreement,  however.

A non-perturbative study of scaling ~\cite{ekr} from a thermodynamic
approach  gave values of $b=-0.30$ and $f=0.66$ at $\beta=2.4$, although
systematic errors are not easy to quantify since an ansatz for the
functional dependence was  assumed.

We conclude that the generalised $\beta$-functions  obtained from  sum
rules using differences of interquark potentials with $R \ge 1$ at
$\beta=2.4$ are consistent with each other, so confirming  that these
observables have a common $a$-dependence. The values we obtain 
are $b=-0.35(2)$ and $f=0.61(3)$. This $a$-dependence is not
that obtained by the first two terms in the perturbative expression.

\section{Excited gluonic potentials}

We measure excited potentials for static sources in the E$_u$
representation  of the  group D$_{4h}$ which is the lattice
symmetry~\cite{hyb} of the potential with separation $R$ along a lattice
axis. We used operators which are U-shaped (ie with links joining the
sources of shape $\, \sqcap - \sqcup$) made of fuzzed links(with 13
iterations) with transverse extent one and two lattice units. A
variational method was used to find the  optimum combination for the
lowest energy state in the E$_u$ representation. We find the  lowest
contamination ($h^2=0.018$ at $T=2$) for $R=4$. This implies that the
study of the colour flux distribution  in this case will have a
contamination of order 25\% if $T=3$ is  used in the plaquette
correlation approach as discussed above.

We consider the difference of the symmetric potential (A$_{1g}$
representation  of D$_{4h}$) at $R=2$ and this E$_u$ potential at $R=4$.
This difference  will cause the self-energy terms to cancel.  The result
for  this potential difference  in lattice units is
$V_{E_u}(4)-V_{A_{1g}}(2)=0.754(2)$ yielding, with $T=3$:  $b=-0.54(6)$
and $f=0.68(5)$. Results for $T > 3$ are consistent with those for $T=3$
but have very large  errors except for the action sum rule where we do
see a significant discrepancy.  These results for the generalised 
$\beta$-functions are less precise than those obtained from differences 
among the A$_{1g}$ potentials (see Table~1) and are not fully consistent.
This  could  arise in part from larger lattice artefacts for this 
particular observable  which would be an interesting conclusion.   This
discrepancy can be attributed, however,  to the larger contamination of
excited states in the E$_u$ case  (as discussed above). One measure of
the size of these excited state contributions  is to compare the $T=2$
result with $T=3$ and we find much larger differences  for E$_u$ than
for A$_{1g}$. From  the $T=4$ data for  the action sum rule (which
controls  the value of $b$) for the E$_u$ case, we also conclude that it
has not achieved a plateau  by $T=3$.

Since we have measured the plaquette expectation values in the gluonic
excited  state (E$_u$), we can explore the spatial distribution of the
colour  field and make comparison with models for the gluonic excitation
such as  string and flux-tube models. Such comparisons have been made 
for the ground state (A$_{1g}$) already~\cite{bali,hay}. Comparison with
such models is most appropriate  at large $R$ where string-like
behaviour is expected. Here we are able to reach $R=8$ which corresponds
to a separation of order 1 fm. Unfortunately, the  contamination from
excited states (here we mean higher energy E$_u$ representation  states)
increases rapidly with $R$.  As an exploratory study, we report our
results using $T=3$ for the  generalised Wilson loop since this is the
largest $T$-value with a reasonable signal.  We specialise to the
midpoint ($r_L=R/2$) to minimise the  effect of self-energy components.
The colour field distributions may then be interpreted as applying to  a
state which is predominantly the required lowest energy E$_u$ state  but
with some contamination from excited states. Results for the transverse
dependence of the colour flux at $R=8$  are shown in  Figure~1 where
they are  compared with the symmetric (A$_{1g}$) potential case. A
qualitatively  similar behaviour is seen for $R=6$.

\begin{figure}[ht] 
\vspace{15cm} 
\includegraphics{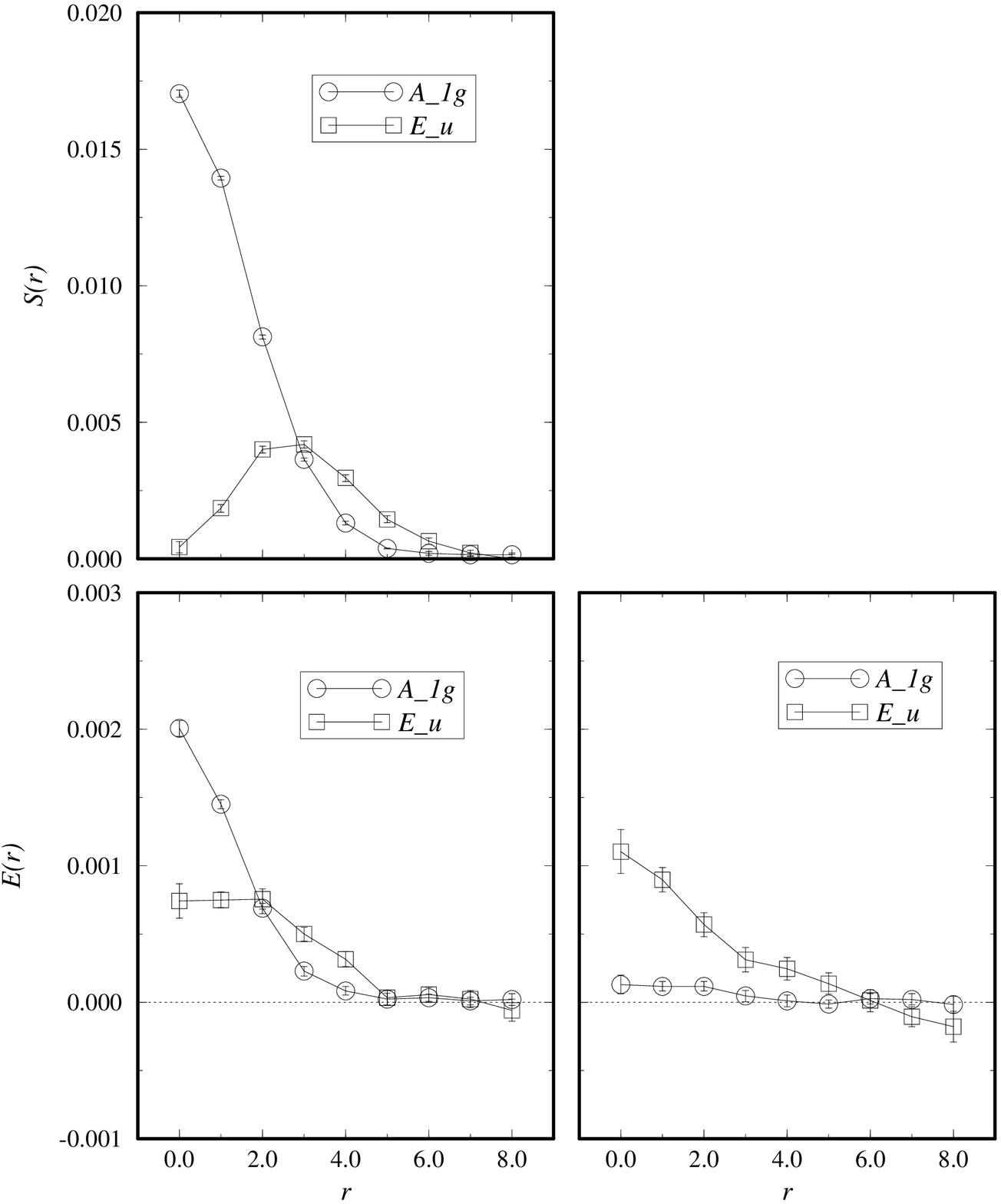} 
\vspace{2cm} 
 \caption{ The colour flux contributions corresponding to the action
($S$), longitudinal ($E_L$, left plot) and transverse energy ($E_T$,
right plot) sum rules of Eqs.~(3--5) for the static quark potential. 
These are shown in lattice units (with $a \approx 0.6$ GeV$^{-1}$)
versus transverse distance $r$  at the mid-point ($r_L=R/2$) for
separation $R=8$.  The data are for the symmetric ground state (A$_{1g}$
representation) and first gluonic excitation (E$_u$ representation).
 } 
\end{figure}

The interpretation of these results is facilitated by the sum rules.
Consider an interquark  potential given by $V(R)=e/R+KR+V_0$, where the
coefficient $e \approx -0.25$ for the A$_{1g}$ potential (this is the
`Coulomb' coefficient) while in string models we expect $e =\pi$ for the
first excited string mode (which  corresponds~\cite{pm} to the E$_u$
case). Then the `transverse energy' sum rule [Eq.~(5)] will relate the
sum over  transverse energy fluctuations to $e$. This explains the much
larger  transverse energy seen in  Figure~1 for the excited gluonic
case.  The other two sum rules suggest that the action and longitudinal
energy will have  fairly similar contributions when summed over $r_T$
for the two cases (the left hand side is $2K$ in each case).  In fact we
see a flatter $r_T$ distribution for the E$_u$ case but the integrals
over $r_T$ are more alike. This flatter distribution in $r_T$
corresponds to a `fatter' flux tube for the excited gluonic case. 
A more detailed discussion will be found in~\cite{big}.

In a string model, such as the Isgur-Paton flux tube model~\cite{IP}, 
the excited modes of the string (which contribute to the E$_u$
potential) have wavefunctions which are suppressed at $r_T=0$. This
gives a `fatter'  distribution. Unfortunately, in the simplest such
picture, this applies only  to the transverse energy while the other two
combinations (action and longitudinal energy) would be little changed.

In conclusion, lattice techniques are able to explore the structure 
of string-like states in both the ground state and excited states. This 
gives support for such a string-like picture in general terms but 
models are not yet successful at reproducing our results in detail.

The authors wish to  acknowledge that these calculations were carried
out at the Centre for Scientific Computing's C94 in Helsinki and the
RAL(UK) CRAY Y-MP and J90.  This work is part of the EC Programme
``Human Capital and Mobility'' -- project number ERB-CHRX-CT92-0051.


\begin{thebibliography} {99}


\bibitem{mich} C. Michael, Nucl. Phys. B280 (1987) 13.

\bibitem{rothe} H. Rothe, Phys. Lett. B355 (1995) 260.

\bibitem{gl} I. H. Jorysz and C. Michael, Nucl. Phys. B302 (1988) 448.

\bibitem{wup} G. Bali, C. Schlichter and  K. Schilling, 
 Phys.Lett. B363  (1995) 196.

\bibitem{cm} C. Michael,  Phys.Rev. D53 (1996) 4102.
  
\bibitem{karsch} F. Karsch, Nucl. Phys. B205 (1982) 285.
  
\bibitem{hyb}              L.A. Griffiths, C. Michael and P.E.L. Rakow,
 Phys. Lett. 129B, 351 (1983)
	   
\bibitem{pm} S.  Perantonis and C. Michael,
Nucl. Phys. B347, 854-68 (1990).

\bibitem{phm} S. Perantonis, A. Huntley and C. Michael, Nucl. Phys.
{ B326}, 544 (1989)

\bibitem{ekr} J. Engels, F. Karsch and K. Redlich, Nucl. Phys. B435
(1995) 295. 

\bibitem{bali} G. Bali, K. Schilling and C. Schlichter, 
Phys. Rev. D51 (1995) 5165.

\bibitem{hay} R.W.~Haymaker, V.~Singh and Y.Peng, 
  Phys.Rev. D53 (1996) 389.
  
\bibitem{IP} N. Isgur and J.E. Paton, Phys. Rev.  D31, 2910 (1985)

\bibitem{big} A. M. Green, C. Michael and P. S. Spencer,  `The Structure
of Flux-tubes in $SU(2)$' in preparation.

\end{thebibliography}
\end{document}